# Effect of plasmonic near field on the emittance of plasmon-enhanced photocathode

Zeng-gong Jiang[1,2], Qiang Gu[1,*], Xu-dong Li[1], Meng Zhang[1], Duan Gu[1]

[1]Shanghai Institute of Applied Physics, Chinese Academy of Sciences, Shanghai 201800, People's Republic of China

[2]University of the Chinese Academy of Sciences, Beijing, 100049, People's Republic of China

guqiang@sinap.ac.cn

**Abstract:**
The introduction of the surface plasmon polarizations makes the emittance of the photocathode complicated. In this paper, the emittance of plasmon-enhanced photocathode is analyzed. It is first demonstrated that the plasmonic near field can increase the emittance of the plasmon-enhanced photocathode. A simulation method has been used to estimate the emittance caused by plasmonic near field, and the suppression method also has been discussed, both of which are significant for the design of high performance plasmon-enhanced photocathode.

## 1. Introduction

With the virtue of prompt response (ps-fs), high brightness, and low emittance etc., the photocathode has becomes a significant enabling technology for the 4[th] generation light sources and advanced accelerator and electron microscopy technology, such as X-ray free electron laser, inverse Compton scattering sources, energy recovery linac, and dynamic transmission electron microscopy etc [1-3]. The metal photocathode has been researched for a long time and used in many facilities, due to its virtue of high brightness, long lifetime, and low cost. And it has been indicated that it holds great promise to eject a uniform 3D ellipsoid electron bunch with only linear internal space charge fields because of its ultrafast response characteristic, which is an ideal distribution for a high-brightness FEL operation [4-6]. But, in order to ensure a high photoemission yield, the metal photocathode usually needs to be pumped by high power ultraviolent laser, which undoubtedly increases the demand and cost for the drive laser [7].

In recent years, the plasmon-enhanced photocathode (PEP) attracts more attentions, which introduces the surface plasmon polarizations (SPPs) to enhance the interaction between laser and materials [8]. With the help of SPPs, this kind of photocathode has the potential to yield much more electrons than traditional metal photocathode [9-11], and it can operate with longer wavelength laser, which reduces the demand of drive laser system [7, 12]. Moreover, it has longer lifetime, lower vacuum requirement, and relatively more mature preparation technology than some high quantum efficiency semiconductor photocathodes.

However, emittance is also a vital parameter for the usefulness of electron sources in many applications [13-18]. When the surface plasmon polarizations (SPPs) are introduced into the photocathode, the emittance of the bunch increases [12], and the mechanism of which becomes complicated and unknown. All of these will restrict the development of high performance plasmon-enhanced photocathode [19].

In this paper, based on the theory of SPPs, we research the influence of SPPs on the emittance of PEP. And a plasmon-enhanced photocathode excited with square hole array has been designed to operate at 532nm. Using this PEP, we demonstrate and research the influence of plasmonic near field on the emittance.

## 2. Characteristics of Plasmon-enhanced photocathode

The process of photoemission can be broken up into three steps: (1) photon absorption and conversion between photons and excited electrons, (2) electron transport to the surface, and (3) crossing the metal-vacuum barrier. When the SPPs are excited, the couple between the electron in the conductor and the pumping electromagnetic wave will be enhanced, contributing to a dramatic increase in the metal's absorption. Moreover, the electric magnetic field near the interface will redistribute, presenting field localization and enhancement. This will improve the energy conversion and reduce the travel distance of the excited electrons. So the introduction of the SPPs can effectively enhance the photoemission of metal photocathode.

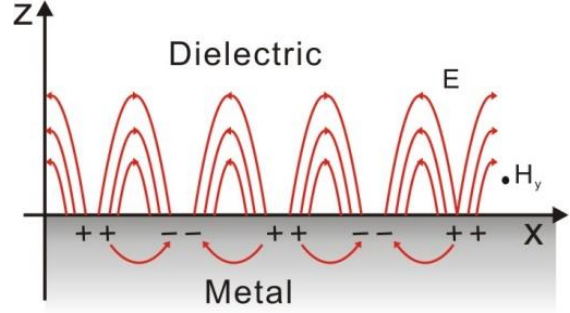

Figure. 1 The schematic diagram of surface plasmon polarizations[20].

In photoinjector, the cathode is usually powered by a laser beam of plane wave, in which the electrons emitted by the cathode can gain no net energy (Lawson-Woodward) [21]. However, with the introduction of SPPs, the electromagnetic field near the dielectric-vacuum interface will be distributed, and an enhanced plasmonic near field (PNF) will appear and decay exponentially perpendicular to the interface [22, 23]. Take a flat interface for example, as shown in Fig. 1, the propagating wave solutions can be given as Formulations (1)-(3) by solving Maxwell's equations.

$$H_y(z) = A e^{i\beta x} e^{-k_1 z} \quad (z>0) \tag{1}$$

$$E_x(z) = iA \frac{1}{\omega \varepsilon_m \varepsilon_d} k_2 e^{i\beta x} e^{-k_1 z} \quad (z>0) \tag{2}$$

$$E_z(z) = -A \frac{\beta}{\omega \varepsilon_m \varepsilon_d} e^{i\beta x} e^{-k_1 z} \quad (z>0) \tag{3}$$

Where A is a constant, $\varepsilon_m$ and $\varepsilon_d$ are the dielectric constants of the metal and the dielectric respectively, $\omega$ is the frequency of the pumping electromagnetic wave, $\beta = k_x$ is the wave vector in the direction of propagation, $k_1$ is the wave vector perpendicular to the interface in the media of the dielectric. Based on the electric magnetic wave solutions (1)-(3), the plasmonic near field contains the electric field component $E_x$ and magnetic field component $H_y$, presenting periodic distribution along the interface with the wavelength of $\frac{2\pi}{\beta}$. Because every electron emitted from PEP need to go through this area, it maybe bring in extra transverse momentums to the electrons, degrading the quality of the beam.

## 3. Demonstration of the emiitance increasing effect of PNF

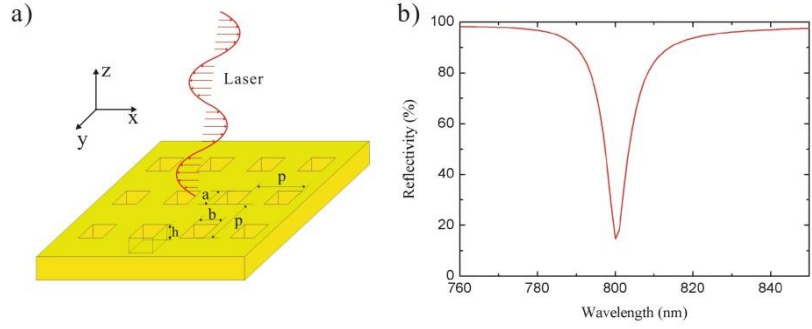

Figure 2. The structural diagram of the Plasmon-enhanced photocathode (a) and its SPPs excitation spectrum (b).

In order to demonstrate whether the surface polarization field of the plasmon can influence the emittance of the bunch, we take a periodic square hole array to excite SPPs and use the CST Studio to simulate it. Figure 2 a) shows the structural diagram of the photocathode which made of Au, the structural parameters are shown as following, a =b=240 nm, h=300 nm, p=755 nm. When the incident laser of 800 nm is irradiated on the surface of the PEP with p polarization perpendicularly, the SPPs are excited successfully, and the photocathode presents an abnormally enhanced absorption of the incident laser which is close to 90%, as shown in Fig. 2 b). With the excitation of SPPs, the plasmonic near field especially close to the surface of photocathode presents a periodic distribution, and aligns with the periodic structure of the photocathode. Figure 3 is the simulation results of the plasmonic near field distribution. Figure 3 a) is the electric field component distribution of PNF, Fig. 3 b) is its magnetic field component distribution. Because of the p polarization incident laser and its electric field component oscillating in the XOZ plane, the plasmonic near field does not contain the electric field component in Y direction and the magnetic field component in X direction. So if the plasmonic near field has an effect on the transverse momentum of the bunch, it maybe only influence the velocity of the electron in X direction, that is, it will only influence the transverse phase-space distribution in X direction.

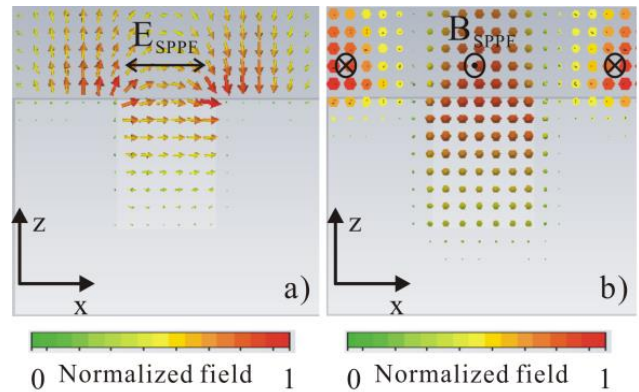

Figure 3. The plasmonic near field distribution of single cycle structure of the array: a) distribution of electric field components, b) distribution of magnetic field components.

The transverse phase-space distributions of the bunch shown in Fig. 4 are simulated by the Tracking module of the CST Particle Studio. In the simulation, we assume the electrons are emitted from the entire surface uniformly without any initial energy. The electrons emitted from the photocathode are accelerated by a 10 MV/m launching field directly. Figure 4 a) and b) respectively present the transverse phase-space distributions of the bunch at X and Y direction when the SPPs are not excited. Due to the symmetrical cathode structure and static electric field distribution, the transverse phase-space distributions at X and Y are the same. The spikes in the transverse phase-space result from the distorted static electric field, which is caused by the periodic holes. When the plasmonic near field is introduced, the transverse phase-space distributions of the bunch at X and Y direction are shown in Figure 4 c) and d), respectively. The plasmonic near field is excited by the incident laser of 0.01 GW/cm$^2$. Compared Fig. 4 d) with b), we can find that, the plasmonic near field has little effect on the transverse phase-space of the bunch at Y direction. But the transverse phase-space at X direction shown in Fig. 4 c) presents cyclical fluctuations, which is obviously different with Fig. 4 a). In the simulations, the electrons always are emitted uniformly and have no initial energy, so the difference of the transverse phase-space distributions is just caused by the plasmonic near field, which demonstrates that the plasmonic near field can increase the emittance of the bunch.

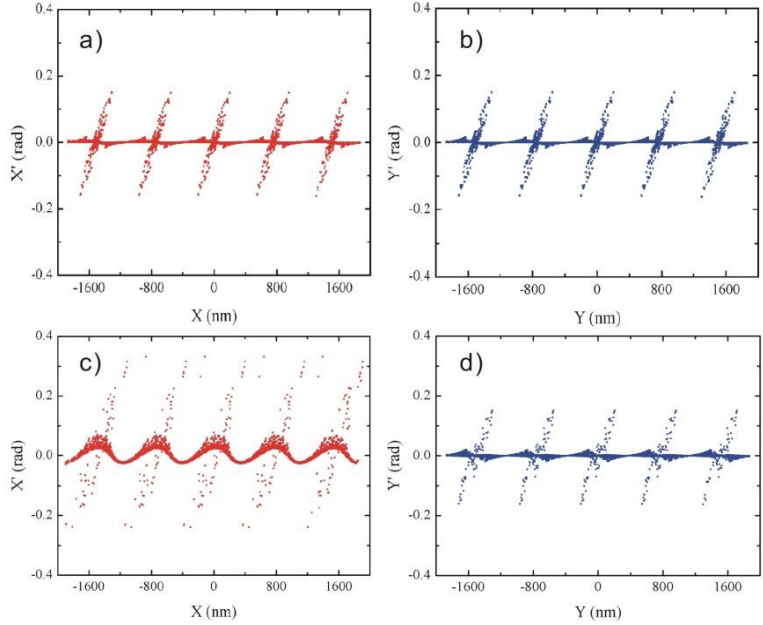

Figure 4. The transverse phase spaces of the bunch in different situations: a) the transverse phase-space distribution at X direction without PNF, b) the transverse phase-space distribution at Y direction without PNF, c) the transverse phase-space distribution at X direction with PNF, d) the transverse phase-space distribution at Y direction with PNF.

**4. Calculation and Analysis of the emittance increased by PNF**

Because the PNF above the surface of photocathode decays with an exponential way, its influence on the beam quality will just occur in a limited area close to the surface of the cathode. So it is not only related with the characteristics of plasmonic near field, but also connected with the movement characteristics of the electrons in this area which are determined by the launching electric field, roughness, and electrons' initial velocities, etc. The emittance growth factor can be given by doing statics of the phase spaces with and without plasmonic near field, which is shown as formula (9).

$$\eta = \frac{\varepsilon_{\text{With SPPs}-X \text{ or } Y}}{\varepsilon_{\text{Without SPPs}}} \tag{9}$$

where $\varepsilon_{\text{With SPPs-X or Y}}$ is the projected emittance of the bunch at X or Y direction with the introduction of PNF, $\varepsilon_{\text{Without SPPs}}$ is the projected emittance of the bunch without PNF. Due to the symmetry, the projected emittance $\varepsilon_{\text{Without SPPs}}$ can be obtained as $\varepsilon_{\text{Without SPPs}} = \varepsilon_{\text{Without SPPs-X}} = \varepsilon_{\text{Without SPPs-Y}}$. Figure 5 presents the emittance growth factors of the square hole array PEP with different launching fields (10-30 MV/m). The plasmonic near fields imported in the simulations are excited by the incident lasers of 0.1-0.6GW/cm$^2$. In the whole simulation, we assume the electrons emitted from the surface of the photocathode are uniform, and their initial energy is constant. Figure 5 shows that the projected emittances at X direction are apparently larger than those at Y direction. The projected emittances at X direction changes with the increase of the incident laser power density, but those at Y direction are almost constant, which are accord with the analysis of Fig. 4 c) and d).

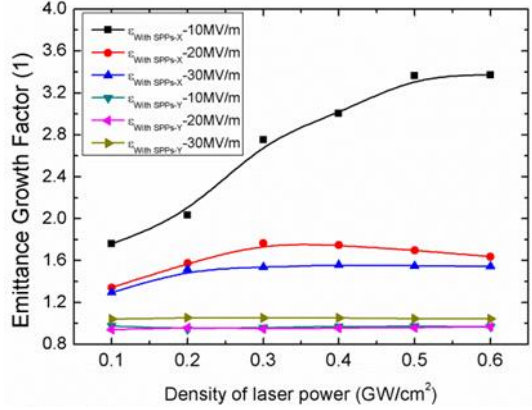

Figure 5. The emittance growth factor of the square hole array PEP.

In order to study the emittance caused by PNF, we need to separate it from the whole emittance. Compared the two simulations, the difference between them is whether introducing the plasmonic near field. For the situation without PNF, the emittance of beam can be written as formula (10), which contains intrinsic thermal emittance and the part caused by roughness of the cathode. When the PNF is introduced, the emittance mainly contains three parts: the intrinsic thermal emittance, the emittance caused by roughness, and the emittance caused by plasmonic near field, which can be given by formula (11).

$$\varepsilon_{\text{Without PNF}} = \sqrt{\varepsilon_{\text{Intrisic}}^2 + \varepsilon_{\text{Roughness}}^2} \tag{10}$$

$$\varepsilon_{\text{With PNF}} = \sqrt{\varepsilon_{\text{Intrisic}}^2 + \varepsilon_{\text{Roughness}}^2 + \varepsilon_{\text{PNF}}^2} \tag{11}$$

Where $\varepsilon_{\text{Roughness}}$ is the emittance caused by the roughness, $\varepsilon_{\text{PNF}}$ is the emittance caused by the introduction of the plasmonic near field. Because the same electron emissions are assumed in the two simulations, both of the intrinsic thermal emittances can be written as $\varepsilon_{\text{Intrisic}}$. Combined the formula (10) and (11), the emittance caused by the plasmonic near field can be obtained by formula (12)

$$\varepsilon_{\text{PNF}} = \sqrt{\varepsilon_{\text{With PNF}}^2 - \varepsilon_{\text{Without PNF}}^2} \tag{12}$$

To evaluate the emittance $\varepsilon_{\text{PNF}}$ under different operating conditions, the ratios $\varepsilon_{\text{PNF}}/\varepsilon_{\text{Without PNF}}$ are calculated and shown in Fig. 6. Compared with the emittance $\varepsilon_{\text{Without PNF}}$, the emittance $\varepsilon_{\text{PNF}}$ caused by plasmonic near field is very large and cannot be neglected especially under a lower launching field, which implies that the introduction of the plasmonic near field is one of the important factors leading to the increase of the emittance of PEP.

From Fig. 6 a), it can be found that the emittance caused by PNF can be modulated by the operating conditions, such as launching field gradient, the density of the laser power, etc. Due to the limited area of the PNF, an increasing launching field gradient can reduce the time that electrons spend in this area. It will weaken the influence of the plasmonic near field on the movement of electrons, reducing the emittance $\varepsilon_{PNF}$, which is found consistent with results in Fig. 6. As the strength of the PNF increases with the incident laser power density, the electrons will spend more time in PNF area and obtain more transverse momentum, resulting in the increase of the emittance $\varepsilon_{PNF}$. But, when the electrons spend longer time than half oscillation period of the field in this area, the electrons' transverse momentum will be decreased by the reversed plasmonic near field. So the emittance $\varepsilon_{PNF}$ presents saturation and even decrease at last, which allows the PEP operating with an acceptable emittance under a high power laser.

Although the increasing launching field can reduce the emittance $\varepsilon_{PNF}$, Figure 6 b) shows it does not decrease linearly with the increase of the launching field gradient. It is because the locally distorted static electric field caused by the micro-structure of the array has transverse components. With the increase of the launching field gradient, the increasing transverse components can prolong the time that electrons spend in the PNF area, enhancing the influence of the PNF on the emittance of the beam. So the size of the micro-structure for the excitation of SPPs should be smaller so that the transverse component of the locally distorted electric field is lower, if we want to suppress the emittance $\varepsilon_{PNF}$ of PEP by operating with a higher gradient launching field.

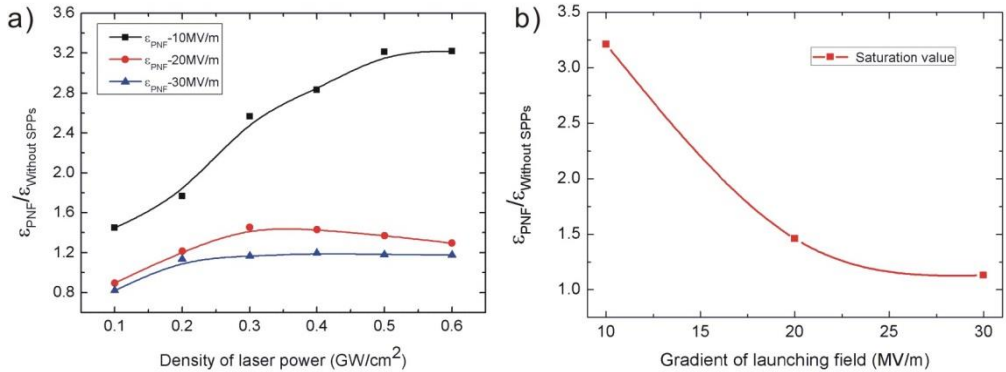

Figure 6. a) The ratio $\varepsilon_{PNF}/\varepsilon_{Without\ PNF}$ of the square hole array PEP under different launching fields and laser power density, b) the saturation value of $\varepsilon_{PNF}/\varepsilon_{Without\ PNF}$ under different launching fields.

## 5. Conclusion

In conclusion, analytical and numerical investigations about the effect of plasmonic near field on the emittance of PEP have been performed. It has been concluded that the introduction of PNF can aggravate the quality of the beam，and is an important factor resulting in the increase of the emittance. A simulation method has been proposed to estimate the emittance $\varepsilon_{PNF}$, which can be extended to optimize the emittances of other PEPs. It has been found that the emittance $\varepsilon_{PNF}$ will increase with the incident laser power density but finally present saturation, which allows the PEP operating under a high power laser and ensures its high yield applications. Moreover, the research denotes that a higher launching field can suppressed the emittance $\varepsilon_{PNF}$ especially when the SPPs are excited with a smaller micro-structure array.

**Acknowledgements**

This work is supported by the National Natural Science Foundation of China (11405251).

**References：**
[1] W. Ackermann, G. Asova, V. Ayvazyan, etc., "Operation of a free-electron laser from the extreme ultraviolet to the water window," Nat. Photonics **1**, 336 (2007).
[2] W. S. Graves, F. X. Kartner, D. E. Moncton, and P. Piot, "Intense superradiant X rays from a compact source using a nanocathode array and emittance exchange," Phys. Rev. Lett. **108**, 263904 (2012).
[3] N.D. Browning, M.A. Bonds, G.H. Campbell, J.E. Evans, T. LaGrange, K.L. Jungjohann, D.J. Masiel, J. McKeown, S. Mehraeen, B.W. Reed, M. Santala, "Recent developments in dynamic transmission electron microscopy," Curr. Opin. Solid St. M. 16, 23-30 (2012).
[4] O. J. Luiten, S. B. van der Geer, M. J. de Loos, F. B. Kiewiet, and M. J. van der Wiel, "How to Realize Uniform Three-Dimensional Ellipsoidal Electron Bunches," Phys. Rev. Lett. 93, 094802 (2004).
[5] P. Musumeci, J. T. Moody, R. J. England, J. B. Rosenzweig, and T. Tran, "Experimental Generation and Characterization of Uniformly Filled Ellipsoidal Electron-Beam Distributions," Phys. Rev. Lett. 100, 244801 (2008).
[6] Aleksandr Polyakov, "Plasmon Enhanced Photoemission," Ph.D. Thesis, University of California Berkeley, California 94720 (2012).
[7] D.H. Dowell et al, "Cathode R&D for future light sources," Nucl. Instrum. Methods Phys. Res., Sect. A **622**, 685–697 (2010).
[8] J. Le Perchec, P. Queḿerais, A. Barbara, and T. Loṕez-Rı́os, "Why Metallic Surfaces with Grooves a Few Nanometers Deep and Wide May Strongly Absorb Visible Light," Phys. Rev. Lett. **100**, 066408 (2008).
[9] A. Polyakov, C. Senft, K. F. Thompson, J. Feng, S. Cabrini, P. J. Schuck, and H. A. Padmore, "Plamon-Enhanced Photocathode for High Brightness and High Repetition Rate X-Ray Sources ," Phys. Rev. Lett. **110**, 076802 (2013).
[10] Y. Gong, Alan G. Joly, L.M. Kong, Patrick Z. El-Khoury, and Wayne P. Hess, "High-Brightness Plasmon-Enhanced Nanostructured Gold Photoemitter," Phys. Rev. Appl. **2**, 064012 (2014).
[11] Subramanian Vilayurganapathy, Manjula I. Nandasiri, Alan G. Joly, Patrick Z. El-Khoury, Tamas Varga, Greg Coffey, Birgit Schwenzer, Archana Pandey, Asghar Kayani, Wayne P. Hess, and Suntharampillai Thevuthasan, "Silver nanorod arrays for photocathode applications," Appl. Phys. Lett. **103**, 161112 (2013).
[12] R. K. Li, H. To, G. Andonian, J. Feng, A. Polyakov, C. M. Scoby, K. Thompson, W. Wan, H. A. Padmore, and P. Musumeci, "Surface-Plasmon Resonance-Enhanced Multiphoton Emission of High-Brightness Electron Beams from a Nanostructured Copper Cathode," Phys. Rev. Lett. **110**, 074801 (2013).
[13] Patrick G. O'Shea, "Reversible and irreversible emittance growth," Phys. Rev. E. **27** (1), 57-1081 (1998).
[14] P. G. O' Shea, S. C. Bender, D. A. Byrd, J. W. Early, D. W. Feldman, C. M. Fortgang, J. C. Goldstein, B. E. Newnam, R. L. SheSeld, R. W. Warren, and T. J. Zaugg, "Ultraviolet Free-Electron Laser Driven by a High-Brightness 45-MeV Electron Beam," Phys. Rev. Lett. **71** (22), 3661 (1993).

[15] D. Shiffler, J. Luginsland, M. Ruebush, M. Lacour, K. Golby, K. Cartwright, M. Haworth, and T. Spencer, "Emission uniformity and shot-to-shot variation in cold field emission cathodes," IEEE Trans. Plasma Sci. **32**, 1262 (2004).

[16] T. Mayer, D. Adams, and B. Marder, "Field emission characteristics of the scanning tunneling microscope for nanolithography," J. Vac. Sci. Technol. B 14, 2438 (1996).

[17] J. W. Lewellen, "Future directions in electron sources," in Proc. of the 21st Particle Accelerator Conference, Knoxville, TN, 2005 (IEEE, Piscataway, NJ, 2005), pp. 563–567.

[18] A. Pedersen, A. Manolescu, and A. Valfells, "Space-Charge Modulation in Vacuum Microdiodes at THz Frequencies," Phys. Rev. Lett. 104, 175002 (2010).

[19] Kevin L. Jensen, Donald A.Shiffler, John J. Petillo, Zhigang Pan, and John W. Luginsland, "Emittance, surface structure, and electron emission," Phys. Rev. ST Accel. Beams **17**, 043402 (2014).

[20] Oliver Benson, "Assembly of hybrid photonic architectures from nanophotonic constituents," Nature. 480, 193-199 (2011).

[21] Q. Kong, et al., Phys. Plasmas. **10**,12 (2003); R. B. Palmer, Part. Accel. **11**, 81 (1980); R. H. Pantell and M. A. Piestrup, Appl. Phys. Lett. **32**, 781 (1978); A. M. Sessler, Am. J. Phys. **54**, 505 (1986); K. T. McDonald, Phys. Rev. Lett. **80**, 1350 (1998).

[22] Anatoly V. Zayatsa, Igor I. Smolyaninovb, and Alexei A. Maradudinc, "Nano-optics of surface plasmon polaritons," Phys. Rep. **408**, 131–314 (2005).

[23] Stefan A. Maier, Plasmonics: Fundamentals and applications (Springer Science+Business Media LLC, New York, 2007), p. 34.